\documentclass{article}

\usepackage{arxiv}

\usepackage[utf8]{inputenc} 
\usepackage[T1]{fontenc}    
\usepackage{hyperref}       
\usepackage{url}            
\usepackage{booktabs}       
\usepackage{amsfonts}       
\usepackage{nicefrac}       
\usepackage{microtype}      
\usepackage{lipsum}		
\usepackage{graphicx}
\usepackage{multirow}
\usepackage{cite}

\usepackage{amsmath}
\usepackage{chemformula}

\usepackage{epstopdf}
\usepackage{array}
\usepackage{subfigure}
\usepackage{ragged2e}
\usepackage{booktabs,makecell,multirow,tabularx}
\usepackage[ruled,linesnumbered]{algorithm2e}

\title{NAS-PINN: Neural architecture search-guided physics-informed neural network for solving PDEs}

\author{ Yifan Wang, ~Linlin Zhong\thanks{This paper has been accepted by Journal of Computational Physics.} \\
	School of Electrical Engineering\\
	Southeast University\\
	No.2 Sipailou, Nanjing, Jiangsu Province 210096, P. R. China\\
	\texttt{linlin@seu.edu.cn}\\
}

\date{May 15, 2023}

\renewcommand{\headeright}{}
\renewcommand{\undertitle}{A Preprint}

\begin{document}
\maketitle

\begin{abstract}
Physics-informed neural network (PINN) has been a prevalent framework for solving PDEs since proposed. By incorporating the physical information into the neural network through loss functions, it can predict solutions to PDEs in an unsupervised manner. However, the design of the neural network structure basically relies on prior knowledge and experience, which has caused great trouble and high computational overhead. Therefore, we propose a neural architecture search-guided method, namely NAS-PINN, to automatically search the optimum neural architecture for solving certain PDEs. By relaxing the search space into a continuous one and utilizing masks to realize the addition of tensors in different shapes, NAS-PINN can be trained through a bi-level optimization, where the inner loop optimizes the weights and bias of neural networks and the outer loop the architecture parameters. We verify the ability of NAS-PINN by several numerical experiments including Poisson, Burgers, and Advection equations. The characteristics of effective neural architectures for solving different PDEs are summarized, which can be used to guide the design of neural networks in PINN. It is found that more hidden layers do not necessarily mean better performance and sometimes can be harmful. Especially for Poisson and Advection, a shallow neural network with more neurons is more appropriate in PINNs. It is also indicated that for complex problems, neural networks with residual connection can improve the performance of PINNs.
\end{abstract}

\section{INTRODUCTION}
\label{sec:sec1}
\paragraph{}
PDEs are ubiquitous in both theoretical and practical science, including electromagnetism, fluid mechanics, finance and many other science fields \cite{J2004Numerical}. As a result, it naturally brings the problem of solving PDEs, which is complex and cumbersome. Since most PDEs have no analytical solutions, many numerical methods have been proposed to achieve approximative numerical solutions. Traditional numerical methods such as finite difference method, finite element method and finite volume method basically discretize equations and computational domains into meshes and then obtain numerical solutions in discrete forms. Although these traditional methods can attain high precision and have rigorous mathematical proof, the computational cost will increase exponentially with the dimensions of PDEs, leading to the curse of dimensionality.

\paragraph{}
In recent years, the great improvement of deep learning has sparked a research trend of solving PDEs with deep neural networks (DNN). The practice of solving PDEs with DNNs is supported by the universal approximation theorem \cite{K1989Multi}, which states that DNNs can theoretically approximate any continuous function. Generally, there are two mainstream deep learning approaches to solving PDEs: learning of neural operators and using PDEs as constraints \cite{S2022Partial}. The former is represented by a series of works based on DeepONet \cite{L2021Learning} and Fourier Neural Operator (FNO) \cite{Z2021Fourier}. This kind of method requires considerable amount of numerical results as training data and once trained, the model can handle a bunch of PDEs with the same form but different equation parameters. However, the training procedure is completely data-driven, neglecting the physical information governed by PDEs. The high computational cost caused by the training data collection and the network parameter optimization is also a problem.

\paragraph{}
The iconic work of the latter is physics-informed neural network (PINN), which is the research of interest in this paper. PINN, first proposed by Raissi et al. \cite{M2019Physics}, is a framework which embeds the physical information into the neural network by defining an appropriate loss function. The framework leverages the automatic differentiation (AD) \cite{A2018Automatic} feature of deep learning to simplify the computation of partial derivatives and has proven effective in solving PDEs and discovering PDE parameters. Since proposed, PINN has rapidly gained attention from researchers. To solve PDEs in discrete or irregular computational domains, cPINN \cite{A2020Conservative} and XPINN \cite{A2021Extended}, characterized by computational domain decomposition, were proposed. In the original PINN framework, boundary conditions and initial conditions are softly constrained by defined loss functions. To guarantee the constraints, boundary and initial conditions in simple forms can be explicitly encoded into the neural network. With one step further, the penalty-free neural network (PFNN) \cite{H2021PFNN} employs two independent neural networks for Dirichlet and Neuman boundary conditions respectively, successfully encoding boundary conditions in complicated forms into the model. The gradient of loss function is another research area of interest. The gradient-optimized PINN by Li et al. \cite{J2022Gradient} and gradient-enhanced PINN by Yu et al. \cite{J2022Enhanced} both focus on the gradient information. The former aims to smooth the gradient distribution for faster convergence, and the latter incorporates the gradient information into the loss function for better performance. The AD technique used in PINN, which is fully dependent on values but irrelevant to spatial information, has also attracted much attention. Xiang et al. \cite{Z2022Hybrid} and Chiu et al. \cite{P2022CAN-PINN} introduced spatial information into the model by substituting AD with numerical differentiation (ND) in different forms and successfully ensured that the solution obtained by the model complies with physical laws. Similar works can be found in \cite{R2022Accelerated}, which introduced radial basis function finite difference to replace AD. The refs. \cite{M2021Efficient} , \cite{A2022Rethinking} and \cite{W2022Robust} discussed the sampling strategy of PINN by adjusting the distribution of sampling points and saw a significant improvement.

\paragraph{}
However, few works have been found to look into the design of the neural network structure in PINN. The design of neural networks is a significant topic in deep learning and can substantially affect the performance. Up until now, the neural network in PINN has generally been designed on the basis of prior knowledge and experience and has followed similar routines, which is to construct a neural network with 4 to 6 hidden layers and with the same number of neurons for each hidden layer \cite{M2019Physics, M2022PDEBench}. The relationship between the neural network architecture and the performance of PINN has been studied in several works \cite{L2020Deep, T2023Can, L2022A-PINN}, bet these efforts were still fragmentary and time-consuming.

\paragraph{}
Neural architecture search (NAS) is a kind of algorithm to search for the optimum neural network architecture in a specific search space \cite{B2018Learning}.Traditional NAS algorithms build architectures by permutations of neural network modules, train and test these architectures to determine their performance, and then choose the best neural network architecture based on the performance ranking. Such a discrete process struggles with the problem of low efficiency and high computational costs. Therefore, to reduce computing overhead and to improve search efficiency have been one of the main research focuses in NAS. Progressive neural architecture search (PNAS) is a technique devised by Liu et al. \cite{C2018Progressive} By gradually discarding structures with poor performance during the search phase, PNAS progressively shrinks the search space and successfully improves the efficiency. With the same purpose of reducing the number of parameters and enhancing efficiency, Hieu Pham et al. \cite{Y2020Understanding} proposed efficient neural architecture search (ENAS), which constructs a hypernetwork for parameter sharing. Similar works can be found in \cite{G2018Understanding} and \cite{A2018SMASH}. However, methods based on hypernetwork have still generally dealt with a discrete search space and developed around reinforcement learning (RL) \cite{L1996Reinforcement} technology. Furthermore, Liu et al. \cite{H2018DARTS} led the study of differentiable NAS by proposing differentiable architecture search (DARTS), which loosens up the discrete search space into a continuous one, so that the discrete optimization problem becomes a continuous optimization problem that can be conveniently optimized through gradient-based methods.

\paragraph{}
This paper proposes a neural architecture search-guided physics-informed neural network (NAS-PINN) by incorporating NAS into the framework of PINN. We realize the automatic search of the best neural architecture for solving a given PDE with a modest quantity of data. Masks are utilized for tensors addition to help search for different numbers of neurons in each layer. The effectiveness of the proposed method is demonstrated by numerical experiments on a range of PDEs. By analyzing the numerical results, the characteristics of efficient neural architectures are summarized for guiding the further research on PINNs.

\paragraph{}
The rest of the paper is organized as follows. In Section \ref{sec:sec2}, the general form of PDEs is provided. The method of NAS-PINN is described in detail in Section \ref{sec:sec3}. Section \ref{sec:sec4} presents the numerical results of different PDEs, including Poisson, Burgers and Advection equations. Finally, the work is concluded in Section \ref{sec:sec5}.

\section{PROBLEM STATEMENT}
\label{sec:sec2}
\paragraph{}
The general form of PDEs can be expressed as:

\begin{equation}
\label{equ:equ1}
	\begin{array}{l}
u{(t,{\bf{x}})_t} + {\cal F}({\bf{x}},u(t,{\bf{x}})) = f(t,{\bf{x}}),{\rm{      }}{\bf{x}} \in \Omega ,t \in [0,T],\\
{\cal B}(u) = b(t,{\bf{x}}),{\rm{               }}{\bf{x}} \in \partial \Omega ,\\
{\cal I}(u) = i(t,{\bf{x}}),{\rm{                }}t = 0.
\end{array}
\end{equation}

Where $u(t, \bf{x})$ is the latent solution to be decided, $u(t, \bf{x})_{t}$ is the temporal derivative, ${\cal F}(\cdot)$ is the linear or nonlinear spatial differential operator containing possible orders of spatial derivatives, $f(t, \bf{x})$ is the source term, ${\cal B}(\cdot)$ is the boundary operator calculating boundary values, $b(t, \bf{x})$ is the boundary condition, ${\cal I}(\cdot)$ is the initial operator calculating initial values, $i(t, \bf{x})$ is the initial condition, $\Omega$ is the computational domain and $\partial\Omega$ is the boundary.

\paragraph{}
By applying NAS-PINN, we will solve PDEs in the form above and discuss the characteristics of efficient neural architectures for different PDEs in the following sections.

\section{METHOD}
\label{sec:sec3}
\subsection{Physics-Informed Neural Network (PINN)}
\label{sec:sec3.1}
\begin{figure}
	\centering
	\includegraphics[width=9.5cm]{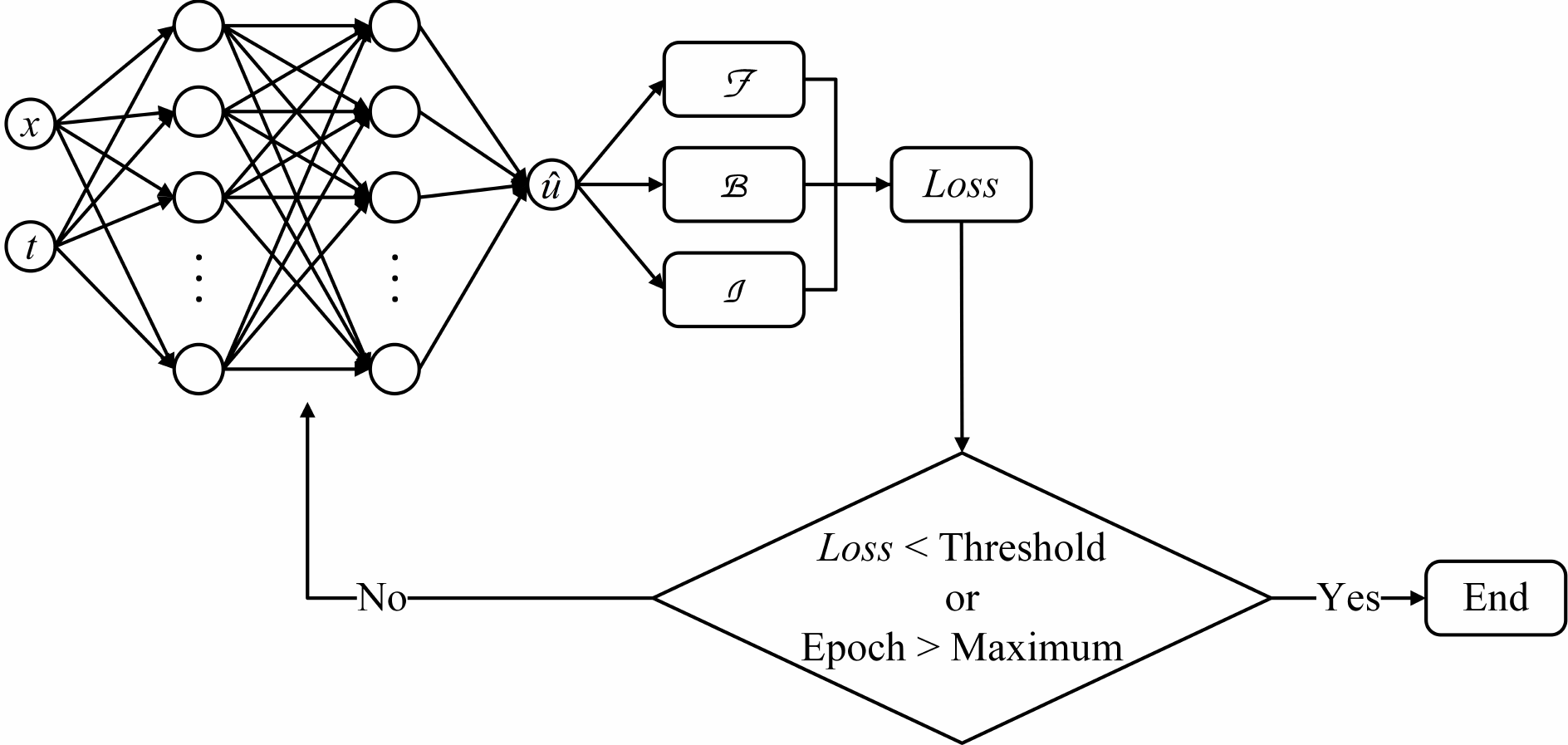}
	\caption{The framework of PINN.}
	\label{fig:fig1}
\end{figure}

\paragraph{}
PINN, first proposed by Raissi et al. \cite{M2019Physics}, is a neural network framework for solving PDEs and discovering PDE parameters by designing an appropriate loss function based on the equation to constrain the network. Here, we focus on the issue of solving PDEs, and a basic framework of PINN is displayed in Figure \ref{fig:fig1}.

\paragraph{}
Considering a PDE in the form of Eq. (\ref{equ:equ1}), the inputs of PINN are spatial coordinates $\bf{x}$ and temporal coordinates $t$ of training points called collocation points and the output is the predicted solution $\hat{u}$. By properly designing the loss function and minimizing it through a certain optimization algorithm, e.g. stochastic gradient descent (SGD), Adam \cite{D2014Adam}, L-BFGS \cite{J1980Updating} and other variants, the output will finally satisfy Eq. (\ref{equ:equ1}) when the network successfully converges. According to Eq. (\ref{equ:equ1}), the loss function can be defined as follows:

\begin{equation}
\label{equ:equ2}
	Loss = {\omega _{\cal F}}{L_{\cal F}} + {\omega _{\cal B}}{L_{\cal B}} + {\omega _{\cal I}}{L_{\cal I}}
\end{equation}

\begin{equation}
\label{equ:equ3}
	{L_{\cal F}} = \frac{1}{{{N_{\cal F}}}}\sum\limits_{i = 1}^{{N_{\cal F}}} {l\left( {\hat u{{({t_i},{{\bf{x}}_i})}_t} + {\cal F}({{\bf{x}}_i},\hat u({t_i},{{\bf{x}}_i})) - f({t_i},{{\bf{x}}_i})} \right)} 
\end{equation}

\begin{equation}
\label{equ:equ4}
	{L_{\cal B}} = \frac{1}{{{N_{\cal B}}}}\sum\limits_{i = 1}^{{N_{\cal B}}} {l\left( {{\cal B}({{\hat u}_i}) - b({t_i},{{\bf{x}}_i})} \right)} 
\end{equation}

\begin{equation}
\label{equ:equ5}
	{L_{\cal I}} = \frac{1}{{{N_{\cal I}}}}\sum\limits_{i = 1}^{{N_{\cal I}}} {l\left( {{\cal I}({{\hat u}_i}) - i({t_i},{{\bf{x}}_i})} \right)} 
\end{equation}

where $\omega_{\cal F}$, $\omega_{\cal B}$ and $\omega_{\cal I}$ are the weighting facotrs for different parts of loss function, $N_{\cal F}$, $N_{\cal B}$ and $N_{\cal I}$ are the numbers of collocation points in the computational domain, the boundary and the initial domain respectively, and $l(\cdot)$ is a certain metric function, which is usually selected as $L^2$ norm or its variants.

\paragraph{}
The PINN framework has been proved to be able to solve PDEs in a variety of circumstances, yet the design of the neural network has not received enough attention. For the remainder of this paper, we will focus on the neural architecture by leveraging NAS.

\subsection{Differentiable NAS}
\label{sec:sec3.2}
\paragraph{}
The number of neural network layers is usually fixed in conventional NAS algorithms, and specific selections of operations are provided for each layer. Such configuration makes the search space discontinuous, as a result of which, it cannot be optimized through gradient-based methods, greatly restricting the convergence speed and efficiency of the algorithm \cite{C2017Model}.

\paragraph{}
Liu et al. \cite{H2018DARTS} proposed DARTS and introduced the concept of differentiable NAS. Let $O$ be a set composed of candidate operations, any of which represents a certain function $o(x)$ for the input $x$. By applying a relaxation to candidate operations, the search space can be made continuous:

\begin{equation}
\label{equ:equ6}
	{\bar o^{(i,j)}}(x) = \sum\limits_{o \in O} {\frac{{\exp \left( {\alpha _o^{(i,j)}} \right)}}{{\sum\nolimits_{o' \in O} {\exp \left( {\alpha _{o'}^{(i,j)}} \right)} }}o(x)} 
\end{equation}

Where ${\bar o^{(i,j}}(x)$ is the mixed operation between the $i$-th layer and the $j$-th layer after relaxation, ${\alpha _o^{(i,j)}}$ is the weight of operation $o$. The discrete process of testing and comparing all possible operation combinations now can be simplified as learning a set of suitable weights ${\alpha _o^{(i,j)}}$ by a gradient-based optimization method. When the algorithm converges, the relaxed search space can be extracted into a discrete neural architecture by selecting the candidate operation with the highest weight.

\paragraph{}
Since the framework above and the basic idea of NAS were first proposed in the field of computer vision, they mostly concentrate on convolutional neural networks, which consists of convolutional layers with different kernel sizes and different pooling layers \cite{B2018Learning}. However, in the context of PINN, applying dense neural networks (DNNs) is the common practice. Therefore, our primary goal in this work is to determine the architecture of DNNs in PINNs, i.e. the number of layers and the neurons in each layer.

\subsection{Masks}
\label{sec:sec3.3}
\begin{figure}
	\centering
	\includegraphics[width=9.5cm]{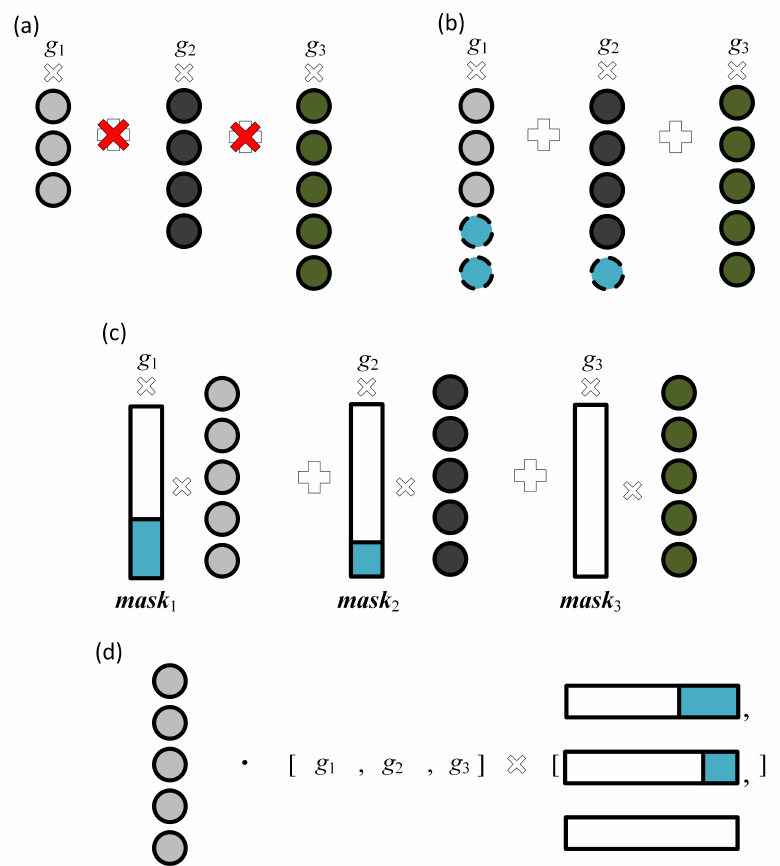}
	\caption{Masks for searching number of neurons. (a) Tensors of different shapes cannot be added together. (b) Padding tensors with zero to yield the tensors of same shape. (c) Equivalent transformation by using one-zero tensor masks. (d) Distributive law of multiplication by sharing weights.}
	\label{fig:fig2}
\end{figure}

\paragraph{}
Although Eq. (\ref{equ:equ6}) reduces the search space into a continuous one, tensor operations only allow tensors of the same shape to be added, making the search for number of neurons impractical, as shown in Figure \ref{fig:fig2}(a). Inspired by the zero-padding in convolutional neural networks, we can pad neurons to the maximum number $k$ \cite{A2020Fbnetv2}, as shown in Figure \ref{fig:fig2}(b). In Figure \ref{fig:fig2}(c), by multiplying the padded neurons by one-zero tensor masks, we deactivate the extra neurons to simulate different numbers of neurons. Finally, by sharing weights, the optional hidden layers can be reduced to one and the output $\bf{y}$ can be expressed as:

\begin{equation}
\label{equ:equ7}
	{\bf{y}} = \sigma \left( {{\bf{w}} \cdot {\bf{x}} + {\bf{b}}} \right) \cdot {\left( {\left[ {{g_1},{g_2},{g_3}} \right] \times \left[ \begin{array}{l}
{\bf{mas}}{{\bf{k}}_1}\\
{\bf{mas}}{{\bf{k}}_2}\\
{\bf{mas}}{{\bf{k}}_3}
\end{array} \right]} \right)^T}
\end{equation}

Where $\sigma(\cdot)$ is the activation function, $\bf{w}$ and $\bf{b}$ are the weights and bias of one single hidden layer, $g_{i}$ is a scalar which is the weight for each number of neurons, $\bf{mask}_{i}$ is the mask for each number of neurons and has a shape of $1 \times k$. Supposing the number of neurons is $j$, then the first $j$-th elements of mask are 1 and the rest $(k-j)$ elements are 0.

\paragraph{}
To determine the number of layers, we introduce the identity transformation as an operation, which represents skipping this layer, and the output $\bf{y}$ becomes:

\begin{equation}
\label{equ:equ8}
	{\bf{y}} = {a_1} \cdot {\bf{x}} + {a_2} \cdot \sigma \left( {{\bf{w}} \cdot {\bf{x}} + {\bf{b}}} \right) \cdot {\left( {\left[ {{g_1},{g_2},{g_3}} \right] \times \left[ \begin{array}{l}
{\bf{mas}}{{\bf{k}}_1}\\
{\bf{mas}}{{\bf{k}}_2}\\
{\bf{mas}}{{\bf{k}}_3}
\end{array} \right]} \right)^T}
\end{equation}

where $a_1$ is the weight for identity transformation which means skipping this layer and $a_2$ is the weight for reserving this layer.

\paragraph{}
Eq. (\ref{equ:equ8}) gives the mapping relations between the input and output of each layer, and by repeatedly applying it, we can build a DNN model in which the most suitable layers can be selected according to weight $a$ and the most suitable neurons in each layer can be decided by weight $g$. Here, we collectively call $a$ and $g$ as $\alpha$, $\bf{w}$ and $\bf{b}$ as $\bf{\theta}$.

\subsection{NAS-PINN}
\label{sec:sec3.4}
\begin{figure}
	\centering
	\includegraphics[width=9.5cm]{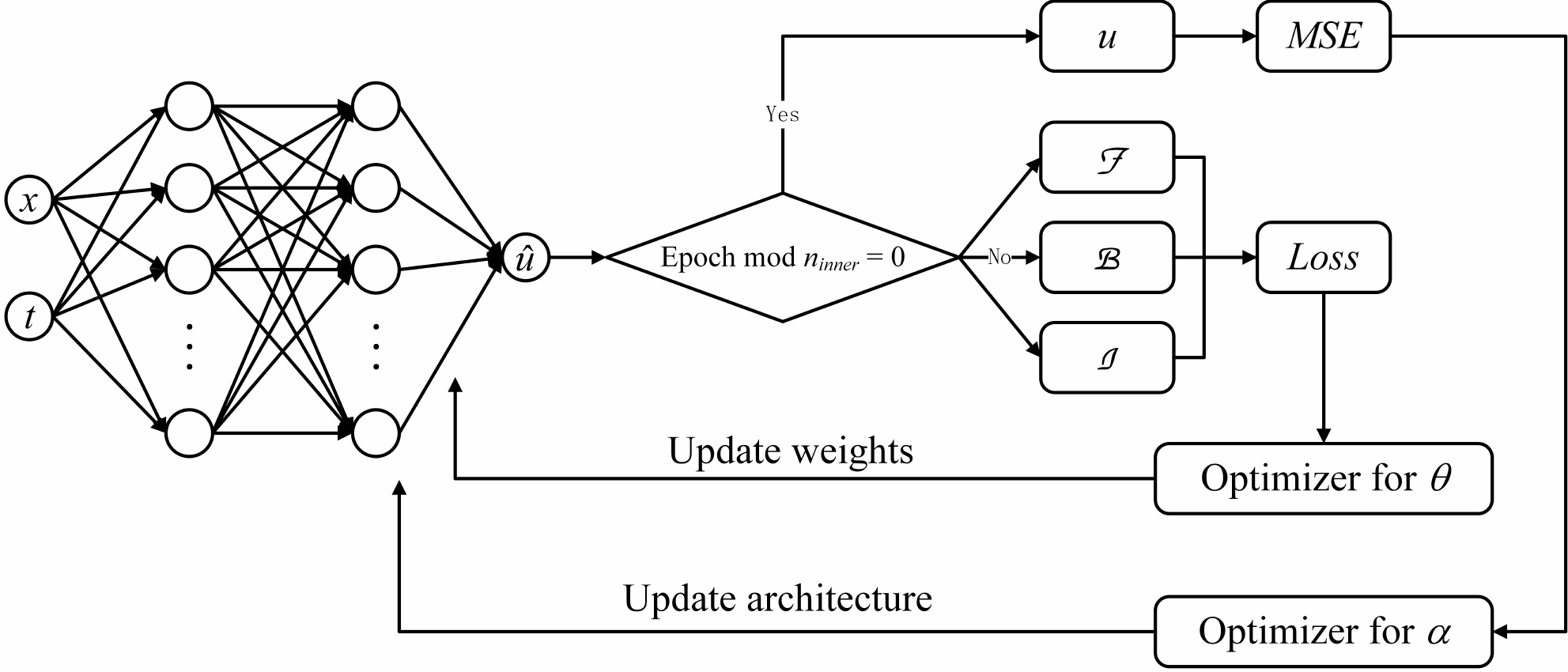}
	\caption{The framework of NAS-PINN.}
	\label{fig:fig3}
\end{figure}

\paragraph{}
Now we can have the whole framework of NAS-PINN as shown in Figure \ref{fig:fig3}, which can be considered as a bi-level optimization problem. In the inner loop, the weights and bias $\bf{\theta}$ of DNN are optimized, while in the outer loop, the optimization object is to find the best $\alpha$. The process can be expressed as:

\begin{equation}
\label{equ:equ9}
	\begin{array}{l}
\mathop {\min }\limits_\alpha  MSE\left( {{{\bf{\theta }}^*},\alpha } \right)\\
s.t.{\rm{   }}{{\bf{\theta }}^*} = \mathop {arg\min }\limits_{\bf{\theta }} Loss\left( {{\bf{\theta }},\alpha } \right)
\end{array}
\end{equation}

\paragraph{}
The loss function for the inner loop can be designed as Eqs. (\ref{equ:equ2}) - (\ref{equ:equ5}) and the loss function for the outer loop can be written as:

\begin{equation}
\label{equ:equ10}
	MSE = \frac{1}{n}\sum\limits_{i = 1}^n {{{\left( {\hat u - u} \right)}^2}} 
\end{equation}

Where $u$ is the known analytical or numerical solution, $n$ is the number of data points and for the outer loop, the required $n$ can be rather small.

\paragraph{}
Such bi-level optimization problem can be solved through an alternate optimization and the corresponding process is demonstrated in Algorithm 1.

\begin{algorithm}[htbp]
	\caption{NAS-PINN}
		create a DNN whose hidden layers are based on Eq. (\ref{equ:equ8})\\
		\label{code:algorithm1}
		set the number of epochs $n_{outer}$ for the outer loop and the number of inner loops $n_{inner}$ in one outer loop\\
		\textbf{while} $epoch < n_{outer}$ \textbf{do}\\
		~~update $\bf{\theta}$ using $Loss$ expressed in Eq. (\ref{equ:equ2})\\
		~~\textbf{if} $epoch$ mod $n_{inner} == 0$\\
		~~~~update $\alpha$ using $MSE$ expressed in Eq. (\ref{equ:equ10})\\
		~~$epoch = epoch+1$\\
		Derive the discrete neural architecture according to $\alpha$\\
\end{algorithm}

\paragraph{}
When the training ends, a discrete neural network model can be derived according to $\alpha$. Basically, we can first determine whether to skip one certain layer by comparing $a_1$ and $a_2$. If the layer is reserved, we can then decide the number of neurons based on $g$. If a certain layer is skipped, there will be no need to investigate its weights $g$.

\paragraph{}
In some cases where $a_1$ and $a_2$ are relatively closed with each other, we assume that skipping the layer and reserving the layer are equally important and we offer a mixed model. In a mixed model, the layers are the combinations of identity transformations and neural network operations. The number of neurons is decided as the same as a discrete one, so these layers can be expressed as:

\begin{equation}
\label{equ:equ11}
	{\bf{y}} = {a_1} \cdot {\bf{x}} + {a_2} \cdot \sigma \left( {{\bf{w}} \cdot {\bf{x}} + {\bf{b}}} \right) \cdot {\left( {{g_{\max }} \times {\bf{mas}}{{\bf{k}}_{\max }}} \right)^T}
\end{equation}

Where $g_{\max}$ is the maximum one among all weights $g$, and $\bf{mask}_{\max}$ is the one-zero tensor mask corresponding to $g_{\max}$.

\section{NUMERICAL EXPERIMENTS}
\label{sec:sec4}
\paragraph{}
In this section, we consider a range of PDEs to test the proposed NAS-PINN and try to find out the characteristics of efficient neural architectures for solving PDEs. Poisson equation, Burgers equation and Advection equation are considered in this work.

\subsection{Poisson equation}
\label{sec:sec4.1}
\paragraph{}
Poisson equation is a class of basic PDEs describing electromagnetic field and heat field, widely applied in electromagnetism and mechanical engineering, etc. Here, we consider a 2-D Poisson equation with Dirichlet boundary condition:

\begin{equation}
\label{equ:equ12}
	\begin{array}{l}
\Delta \varphi (x,y) =  - 2{\pi ^2}\cos (\pi x)cos(\pi y),{\rm{    }}x,y \in \Omega \\
\varphi (x,y) = \cos (\pi x)\cos (\pi y),{\rm{                }}x,y \in \partial \Omega 
\end{array}
\end{equation}

\paragraph{}
This equation can be analytically solved:

\begin{equation}
\label{equ:equ13}
	\varphi (x,y) = \cos (\pi x)\cos (\pi y)
\end{equation}

\paragraph{}
We first consider the Poisson equation in a square computational domain to verify the effectiveness of the proposed NAS-PINN. We construct a relatively small search space, which is a neural network with up to 5 hidden layers and 30, 50 or 70 neurons in each layer. Every possible neural architecture in the discrete search space is trained and tested respectively, as an approximate example of grid search. Then we use NAS-PINN to search for a neural architecture and investigate whether it is the best one. All the 363 architectures in the discrete search space are trained by Adam with 500 collocation points randomly sampled in the domain and 100 boundary points uniformly distributed on the boundary. For the architecture search phase, 1000 collocation points and 200 boundary points are sampled by the same strategies as before to search for the best neural architecture, and Adam is applied in the architecture search phase as well. The obtained neural architecture is then trained from scratch in the same way as the 363 architectures.

\paragraph{}
To make a more comprehensive comparison, a traditional AutoML method SMAC \cite{M2022SMAC3} is also tested. SMAC is a versatile Bayesian optimization package for hyperparameter optimization and for the discussed problem, hyperparameters to be optimized are number of hidden layers and number of neurons. For SMAC, a rather small research space including 15 different neural architectures is set, where the number of neurons for each hidden layer can only be the same. The same 1000 collocation points and 200 boundary points as in the architecture search phase are used for SMAC.

\begin{figure}
	\centering
	\includegraphics[width=14cm]{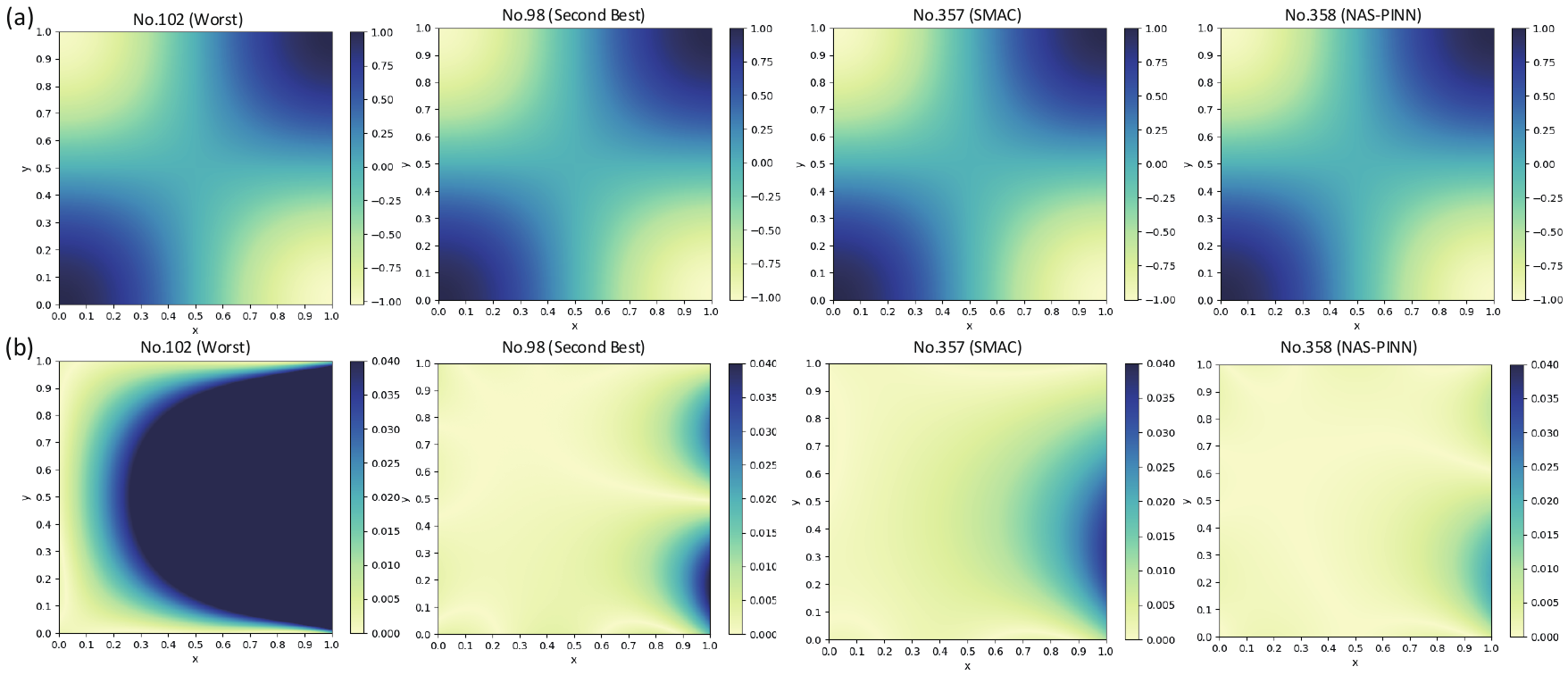}
	\caption{Poisson equation: The predicted solutions (a) and error distributions (b) of different neural architectures.}
	\label{fig:fig4}
\end{figure}

\paragraph{}
Finally, 1,000,000 points are sampled uniformly to test all the converged neural architectures. The predicted solutions and error distributions of different architectures are shown in Figure \ref{fig:fig4} and the $L^2$ error are listed in Table \ref{tab:tab1}. All experiments are repeated 5 times and the $L^2$ error are obtained by the average of the 5 repetitions.

\begin{table}[htbp]
	\centering
	\caption{Poisson equation: $L^2$ error of different architectures}
	\label{tab:tab1}
	\begin{tabular}{|l|l|l|}
		\hline
		Architecture number      & Architecture                   & $L^2$ error  \\
		\hline
		No. 102 (Worst)          & {[}2, 30, 50, 70, 50, 30, 1{]} & $1.97\times10^{-2}$ \\
		\hline	
		No. 98 (Second Best)     & {[}2, 70, 70, 30, 30, 50, 1{]} & $1.57\times10^{-3}$ \\
		\hline
		No. 357 (SMAC)           & {[}2, 50, 50, 1{]}             & $1.95\times10^{-3}$ \\
		\hline
		No. 358 (Best, NAS-PINN) & {[}2, 50, 70, 1{]}             & $4.46\times10^{-4}$ \\
		\hline
	\end{tabular}
\end{table}

\paragraph{}
The architectures are described in the form of sequences in Table \ref{tab:tab1}. The first and the last element of the sequence stand for the input and output channel, while the other elements represent the neuron numbers for each layer. For example, the size of the input for the architecture No.98 is $n\times2$, where $n$ is the batch size, 2 stands for the coordinates $x$ and $y$, and the first hidden layer has 70 neurons. The architecture obtained through NAS-PINN is No.358.

\paragraph{}
From Table \ref{tab:tab1} and Figure \ref{fig:fig4}, we can clearly see that the neural architecture by NAS-PINN has the smallest $L^2$ error and the smallest maximum error value, and its error distribution is improved compared to other architectures as well. Therefore, the proposed NAS-PINN does find out the best neural architecture in the given search space. Besides, although the architecture No.98 also shows relatively good performance (it is the second best architecture in 363 possible architectures), it has much more parameters than the architecture by NAS-PINN, which indicates that more parameters do not necessarily mean better performance and an appropriately designed neural architecture appears to be particularly important. Furthermore, the common sense that a deeper neural network is always better seems not to be true in all circumstances in PINNs. At least for the given Poisson equation, a shallow but wide neural network (a neural network with fewer hidden layers but more neurons in each layer) prevails over the deep ones.

\paragraph{}
Compared with SMAC, NAS-PINN can search in a larger and more flexible search space, so that NAS-PINN is more likely to find the truly best neural architecture. It is also worth pointing out that, although SMAC does the search in a rather small search space with only 15 neural architectures, it takes SMAC 2.08 hours to find architecture No. 357, while NAS-PINN uses 1.57 hours to find architecture No. 358 from 363 different architectures. All the numerical experiments are conducted on Intel (R) Core i9-9900K @ 3.60GHz / NVIDIA GeForce RTX 3090.

\subsection{Poisson equation in irregular computational domains}
\label{sec:sec4.2}
\paragraph{}
To demonstrate the adaptability of NAS-PINN to irregular computational domains, we further inspect the Poisson equation in Eq. (\ref{equ:equ12}) in different computational domains, which include circular, L-shaped and flower-shaped domains. Specifically, the circular domain is a circle with a center at (0.5, 0.5) and a radius of 0.5. The L-shaped domain is the difference set of two squares. The lower left corners of these two squares are at (0, 0) and (1, 1) respectively, while their upper right corners are both at (2, 2).

\begin{figure}
	\centering
	\includegraphics[width=17cm]{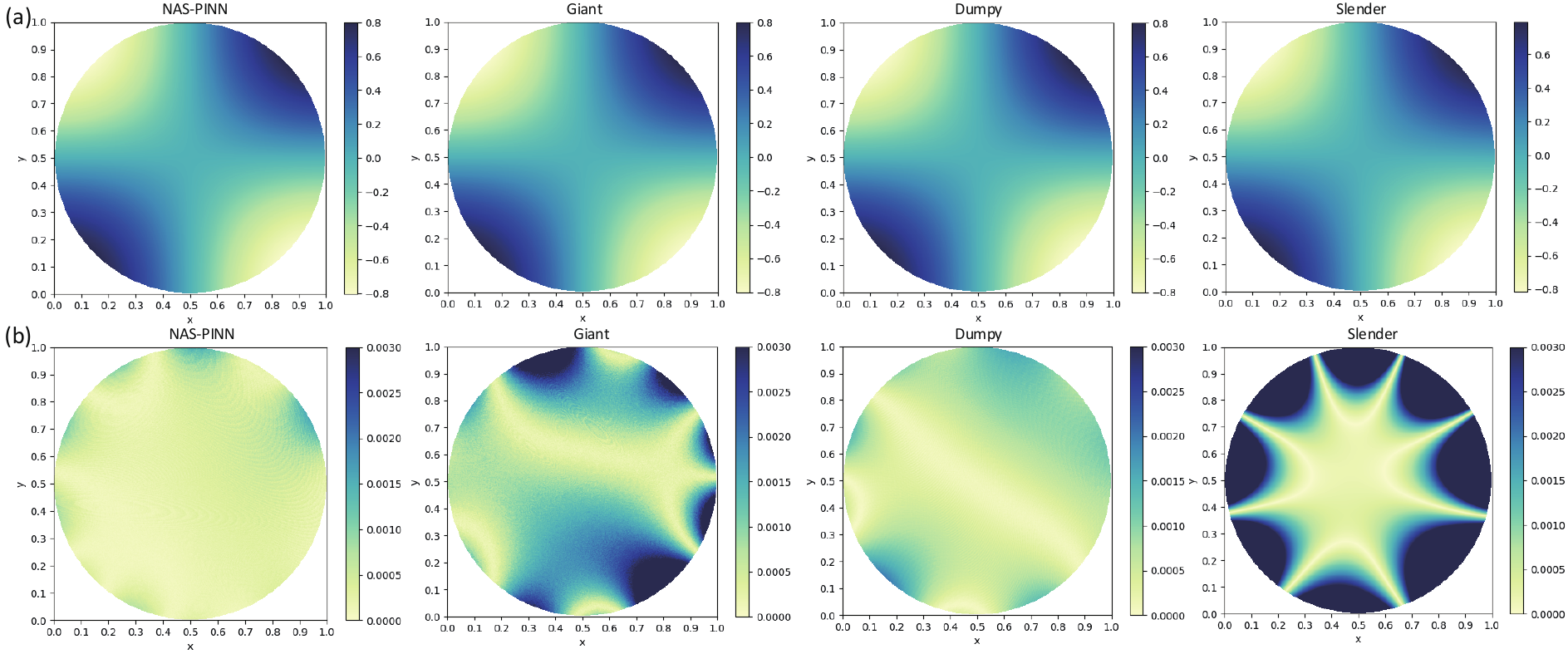}
	\caption{Poisson equation: The predicted solutions (a) and error distributions (b) in the circular computational domain.}
	\label{fig:fig5}
\end{figure}

\begin{figure}
	\centering
	\includegraphics[width=17cm]{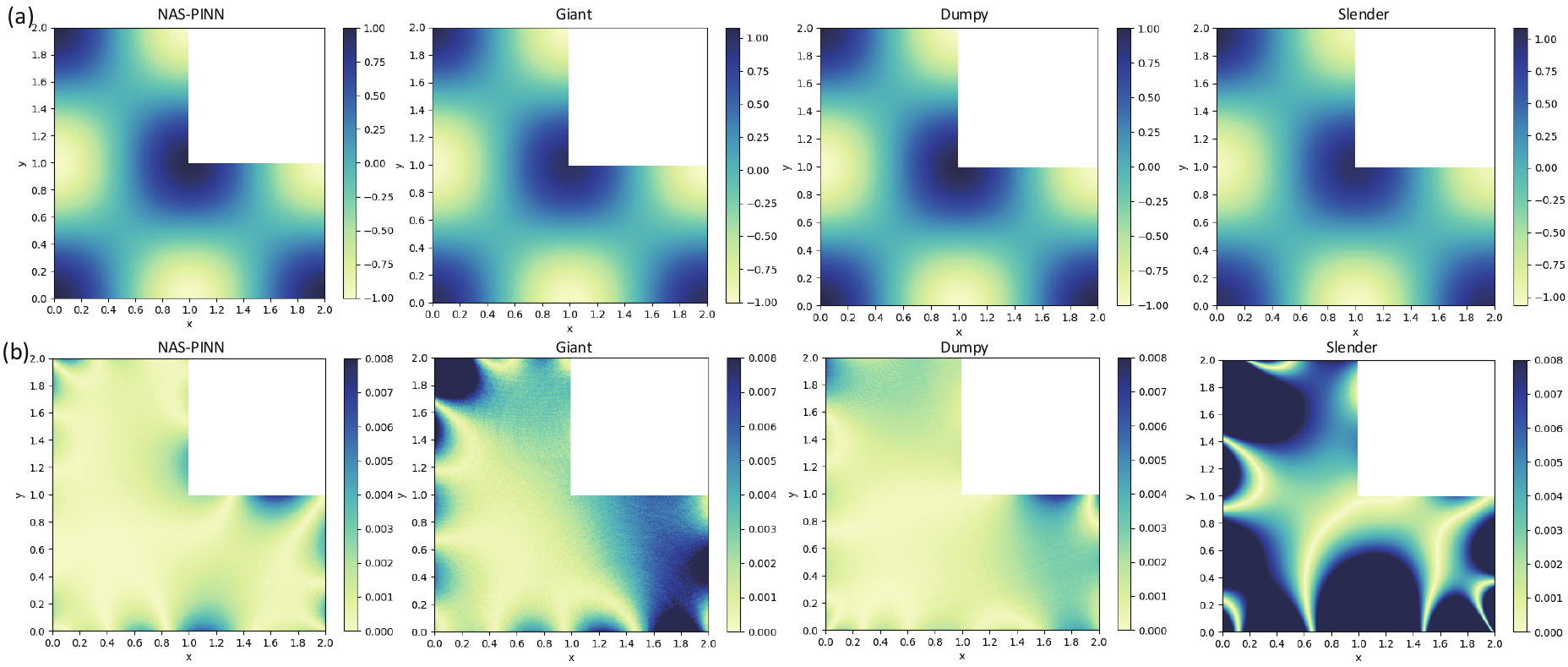}
	\caption{Poisson equation: The predicted solutions (a) and error distributions (b) in the L-shaped computational domain.}
	\label{fig:fig6}
\end{figure}

\begin{figure}
	\centering
	\includegraphics[width=17cm]{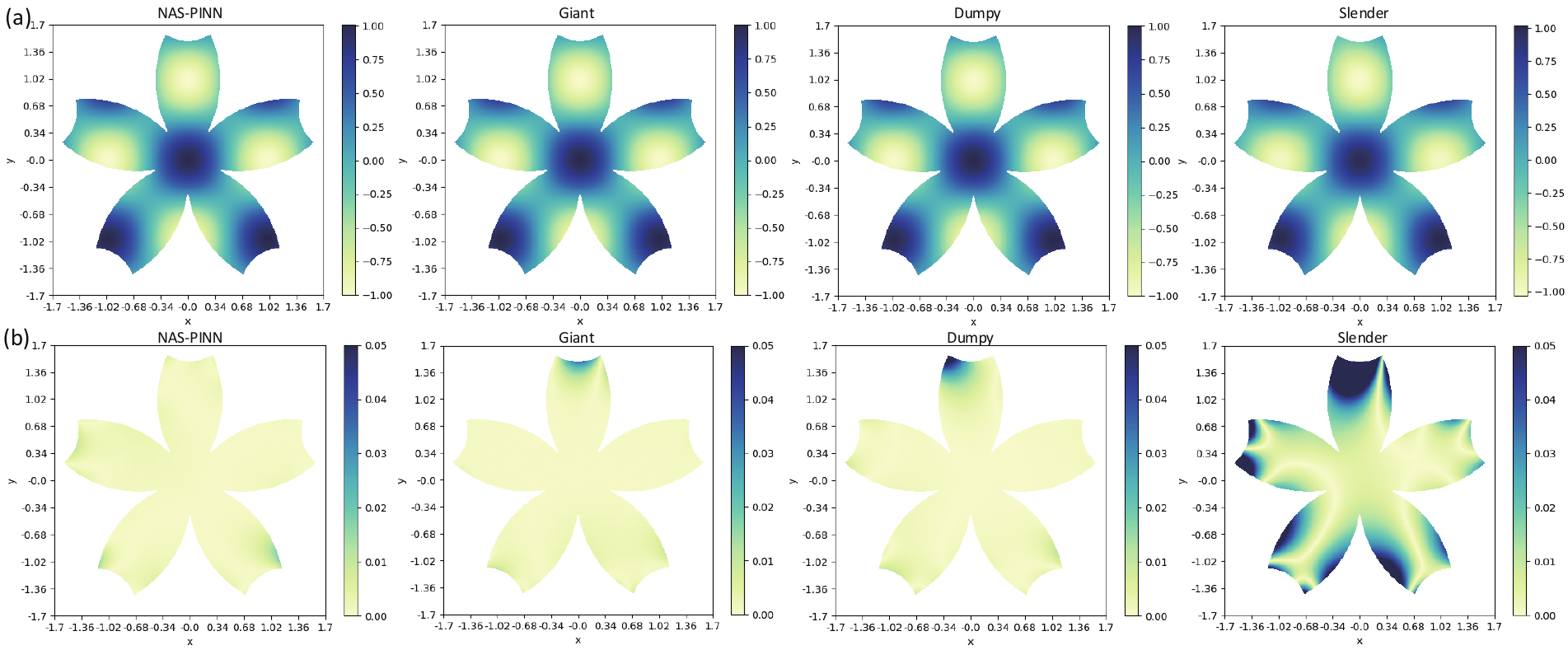}
	\caption{Poisson equation: The predicted solutions (a) and error distributions (b) in the flower-shaped computational domain.}
	\label{fig:fig7}
\end{figure}

\paragraph{}
For irregular computational domains, the search space is set to be a neural network with up to 7 hidden layers, and the numbers of neurons each layer can be selected from 10 to 110 in increments of 20. As irregular computational domains are rather complex, the architecture search phase adopts 2500 collocation points and 500 boundary points. To compare with the neural architecture by NAS-PINN, we manually design three reference neural architectures from the search space based on experience, which are: a) Architecture Giant, a neural network with the most parameters in the search space (7 layers and 110 neurons per layer), b) Architecture Dumpy, a shallow (2 hidden layers) neural network with the most neurons per layer (110 neurons per layer), c) Architecture Slender, a neural network with the most hidden layers (7 layers) but the fewest neurons per layer (10 neurons per layer). In addition to these three basic architectures, one neural architecture with increasing neurons and another with decreasing neurons are investigated, as representation of complex and manually conceivable neural architectures. All the five reference architectures as well as the NAS-PINN-searched architecture are trained from scratch by using 500 collocation points and 100 boundary points. The predicted solutions and error distributions in different computational domains are displayed in Figures \ref{fig:fig5}, \ref{fig:fig6} and \ref{fig:fig7}, and Table \ref{tab:tab2} gives the corresponding $L^2$ error. All the experiments are repeated 5 times and the average values are then calculated.

\paragraph{}
As supposed, irregular computational domains are more challenging to deal with than regular domains, which results in an interesting feature: the weights $a_1$ and $a_2$ are close to each other. We reckon this feature as a preference for residual structures. As Eq. (\ref{equ:equ11}) shows, if $a_1$ and $a_2$ are almost at the same level, the mapping relation becomes a summation of an identity transformation and a neural network operation \cite{K2016Deep}. Therefore, we reserve the residual structures when $a_1$ and $a_2$ are both smaller than a threshold $\varepsilon$, and we call such layers as mixed layers. Mixed layers are expressed in parentheses in Table \ref{tab:tab2}.

\begin{table}[htbp]
	\centering
	\caption{Poisson equation: $L^2$ error in different computational domains}
	\label{tab:tab2}
	\begin{tabular}{|l|l|l|l|}
		\hline
		Computational domain           & Name         & Architecture                          & $L^2$ error  \\
		\hline
		\multirow{6}{*}{Circle}        & NAS-PINN             & {[}2, 110, (50, 50, 50, 30,) 1{]}     & $2.25\times10^{-7}$ \\
		\cline{2-4}
                               		& Giant   & {[}2, $110\times7$, 1{]}                     & $2.55\times10^{-6}$ \\
		\cline{2-4}
                               		& Dumpy   & {[}2, $110\times2$, 1{]}                     & $3.02\times10^{-7}$ \\
		\cline{2-4}
                               		& Slender & {[}2, $10\times7$, 1{]}                      & $1.22\times10^{-5}$ \\
		\cline{2-4}			
							& Increase & {[}2, 10, 30, 50, 70, 90, 110, 110, 1{]}	    & $4.49\times10^{-6}$ \\
		\cline{2-4}
							& Decrease & {[}2, 110, 90, 70, 50, 30, 10, 10, 1{]}	    & $2.55\times10^{-6}$ \\
		\hline
		\multirow{6}{*}{L-shaped}      & NAS-PINN             & {[}2, 110, 110, (10,) 1{]}            & $2.05\times10^{-6}$ \\
		\cline{2-4}
                               		& Giant   & {[}2, $110\times7$, 1{]}                     & $8.39\times10^{-6}$ \\
		\cline{2-4}
                               		& Dumpy   & {[}2, $110\times2$, 1{]}                     & $3.38\times10^{-6}$ \\
		\cline{2-4}
                               		& Slender & {[}2, $10\times7$, 1{]}                      & $4.37\times10^{-4}$ \\
		\cline{2-4}			
							& Increase & {[}2, 10, 30, 50, 70, 90, 110, 110, 1{]}	    & $2.82\times10^{-5}$ \\
		\cline{2-4}
							& Decrease & {[}2, 110, 90, 70, 50, 30, 10, 10, 1{]}	    & $1.62\times10^{-5}$ \\
		\hline
		\multirow{6}{*}{Flower-shaped} & NAS-PINN             & {[}2, 50, 70, (70, 70,) 70, 110, 1{]} & $6.91\times10^{-6}$ \\
		\cline{2-4}
                               		& Giant   & {[}2, $110\times7$, 1{]}                     & $8.97\times10^{-6}$ \\
		\cline{2-4}
                               		& Dumpy   & {[}2, $110\times2$, 1{]}                     & $1.32\times10^{-5}$ \\
		\cline{2-4}
                               		& Slender & {[}2, $10\times7$, 1{]}                      & $1.85\times10^{-3}$ \\
		\cline{2-4}			
							& Increase & {[}2, 10, 30, 50, 70, 90, 110, 110, 1{]}	    & $8.70\times10^{-5}$ \\
		\cline{2-4}
							& Decrease & {[}2, 110, 90, 70, 50, 30, 10, 10, 1{]}	    & $1.31\times10^{-4}$ \\
		\hline
	\end{tabular}
\end{table}

\paragraph{}
The results show that the proposed NAS-PINN can adapt to irregular computational domains well. It is indicated that when the problem is complex, neural architectures with residual connection are more appropriate in PINNs, and the layers with such residual connection tend to appear in the middle and last few layers of the neural networks. Such preference for residual connection is difficult to realize and often neglected when we manually design neural architectures for PINNs. We also notice that Architecture Dumpy performs better than Architecture Giant and Slender in the circular and L-shaped computational domains and none of the NAS-PINN-searched architectures reaches the maximum layers (7 layers). This means a relatively shallow neural network with more neurons is more suitable for solving Poisson equation even in irregular domains. As for the most complex flower-shaped computational domain, NAS-PINN and Giant apparently prevail over other architectures. It is also obvious that the architecture based on NAS-PINN has much fewer parameters than Architecture Dumpy, indicating the capability of NAS-PINN to help design effective architectures and save training cost.

\subsection{Burgers equation}
\label{sec:sec4.3}
\paragraph{}
Burgers equation, as an important part of fluid mechanics and gas dynamics, describes the process of propagation and reflection of shock waves. Here, a time-varying 1-D Burgers equation with periodic boundary condition is considered:

\begin{equation}
\label{equ:equ14}
	\begin{array}{l}
{u_t} + u{u_x} - \left( {\upsilon /\pi } \right){u_{xx}} = 0,{\rm{      }}x \in \left[ { - 1,1} \right],t \in \left[ {0,1} \right]\\
u(0,x) =  - \sin \left( {\pi x} \right)\\
u(t, - 1) = u(t,1) = 0
\end{array}
\end{equation}

Where $\upsilon$ is the diffusion coefficient.

\begin{figure}
	\centering
	\includegraphics[width=17cm]{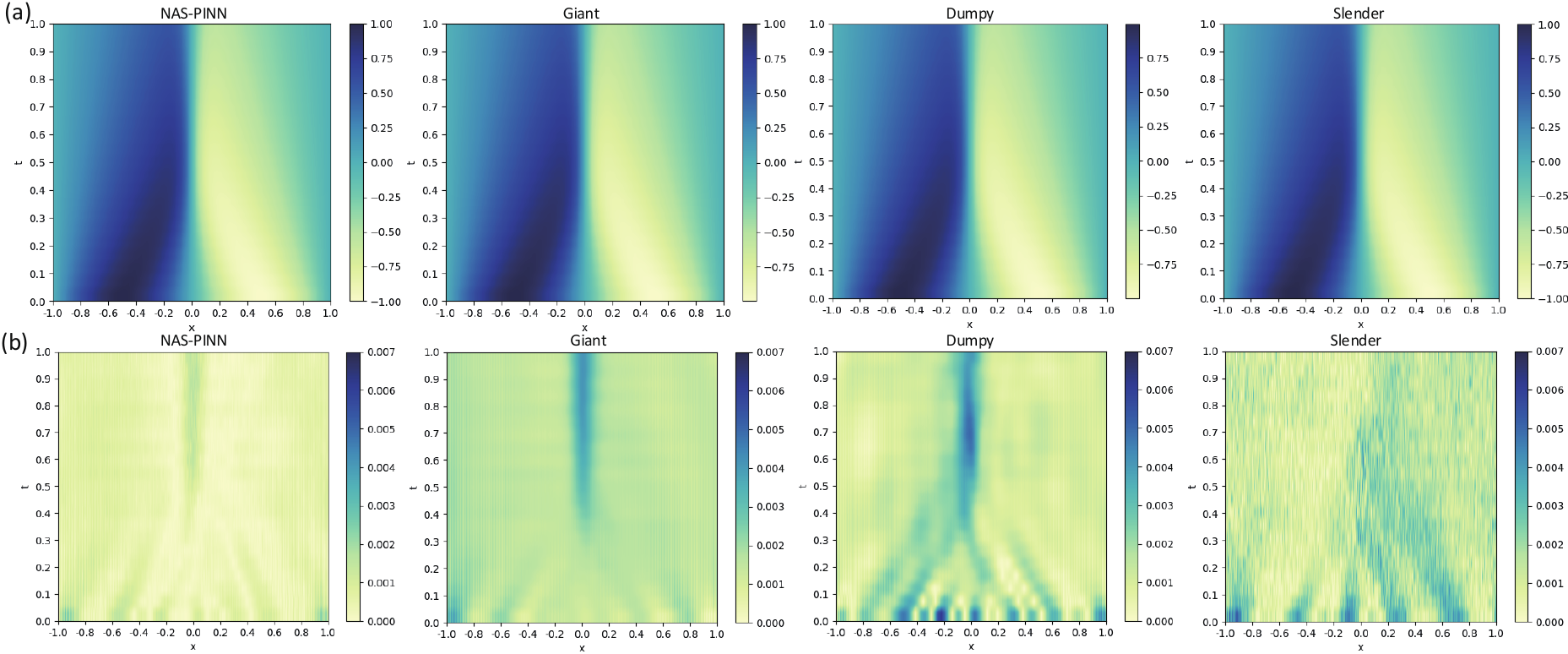}
	\caption{Burgers equation: The predicted solutions (a) and the error distributions (b) of different neural architectures ($\upsilon=0.1$).}
	\label{fig:fig8}
\end{figure}

\begin{figure}
	\centering
	\includegraphics[width=17cm]{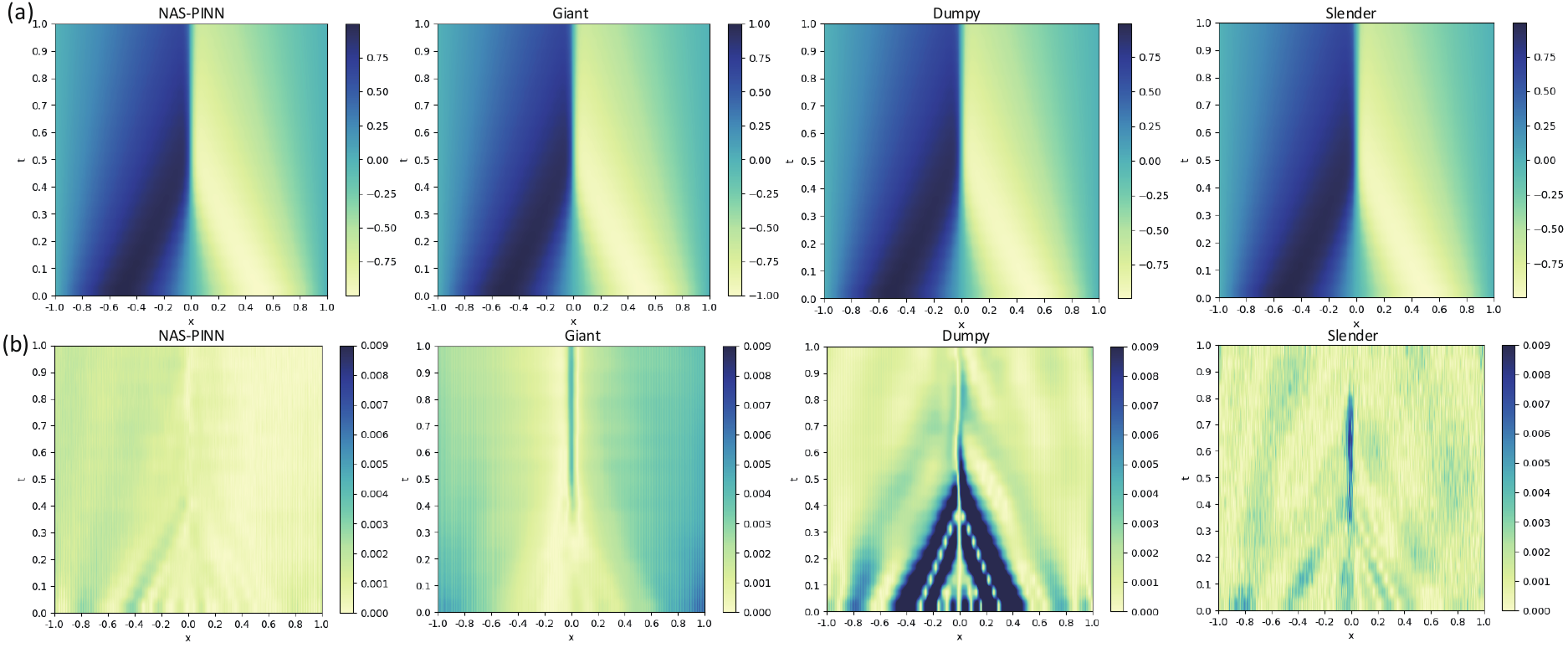}
	\caption{Burgers equation: The predicted solutions (a) and the error distributions (b) of different neural architectures ($\upsilon=0.04$).}
	\label{fig:fig9}
\end{figure}

\paragraph{}
The Burgers equation with different values of $\upsilon$ is investigated and the reference solutions are obtained through Chebyshev spectral method \cite{T2014Chebfun}. The search space and the reference neural architectures are kept the same with those in Section \ref{sec:sec4.2}. In the architecture search phase, we uniformly take 21 points along the $t$-axis and 250 points along the $x$-axis. The same points are used to train all the neural architectures from scratch. 21 points along the t-axis and 500 points along the x-axis are sampled uniformly to test all the converged models. Table \ref{tab:tab3} and Figures \ref{fig:fig8} and \ref{fig:fig9} give the results of Burgers equations with different diffusion coefficients $\upsilon$. The experiments are repeated 5 times and the average values are then calculated.

\paragraph{}
It is found that the predictions of the NAS-PINN-searched architectures are more accurate than those of the reference architectures, and such advantage is especially evident when $\upsilon=0.1$. The NAS-PINN-searched architectures generally have 3 or 4 hidden layers, which is much smaller than the maximum number, further demonstrating that deeper neural networks do not necessarily produce better results, and a best number of hidden layers do exist for a certain problem. Meanwhile, the NAS-PINN-searched architectures prefer to use different numbers of neurons for each hidden layer, which is commonly neglected in experience-based neural network design. By comparing with the reference architectures, it also indicates that for Burgers equations, more hidden layers may be more critical than more neurons, as the architecture Slender can acquire equivalent or even better results compared to the architecture Giant.

\begin{table}[htbp]
	\centering
	\caption{Burgers equation: $L^2$ error with different diffusion coefficients $\upsilon$}
	\label{tab:tab3}
	\begin{tabular}{|l|l|l|l|}
		\hline
		$\upsilon$                       & Name         & Architecture                  & $L^2$ error  \\
		\hline
		\multirow{6}{*}{$\upsilon=0.1$}  & NAS-PINN             & {[}2, 90, 50, 110, 1{]}       & $8.87\times10^{-7}$ \\
		\cline{2-4}
                        		& Giant   & {[}2, $110\times7$, 1{]}             & $1.44\times10^{-6}$ \\
		\cline{2-4}
                        		& Dumpy   & {[}2, $110\times2$, 1{]}             & $1.52\times10^{-6}$ \\
		\cline{2-4}
                        		& Slender & {[}2, $10\times7$, 1{]}              & $1.87\times10^{-6}$ \\
		\cline{2-4}			
							& Increase & {[}2, 10, 30, 50, 70, 90, 110, 110, 1{]}	    & $2.88\times10^{-6}$ \\
		\cline{2-4}
							& Decrease & {[}2, 110, 90, 70, 50, 30, 10, 10, 1{]}	    & $8.97\times10^{-7}$ \\
		\hline
		\multirow{6}{*}{$\upsilon=0.07$} & NAS-PINN             & {[}2, 90, 70, 30, 110, 1{]}   & $1.41\times10^{-6}$ \\
		\cline{2-4}
                        		& Giant   & {[}2, $110\times7$, 1{]}             & $2.15\times10^{-6}$ \\
		\cline{2-4}
                        		& Dumpy   & {[}2, $110\times2$, 1{]}             & $4.94\times10^{-6}$ \\
		\cline{2-4}
                        		& Slender & {[}2, $10\times7$, 1{]}              & $2.35\times10^{-6}$ \\
		\cline{2-4}			
							& Increase & {[}2, 10, 30, 50, 70, 90, 110, 110, 1{]}	    & $4.45\times10^{-6}$ \\
		\cline{2-4}
							& Decrease & {[}2, 110, 90, 70, 50, 30, 10, 10, 1{]}	    & $4.10\times10^{-6}$ \\
		\hline
		\multirow{6}{*}{$\upsilon=0.04$} & NAS-PINN             & {[}2, 110, 110, 70, 110, 1{]} & $1.51\times10^{-6}$ \\
		\cline{2-4}
                        		& Giant   & {[}2, $110\times7$, 1{]}             & $2.70\times10^{-6}$ \\
		\cline{2-4}
                        		& Dumpy   & {[}2, $110\times2$, 1{]}             & $1.60\times10^{-5}$ \\
		\cline{2-4}
                        		& Slender & {[}2, $10\times7$, 1{]}              & $2.61\times10^{-6}$ \\
		\cline{2-4}			
							& Increase & {[}2, 10, 30, 50, 70, 90, 110, 110, 1{]}	    & $3.46\times10^{-6}$ \\
		\cline{2-4}
							& Decrease & {[}2, 110, 90, 70, 50, 30, 10, 10, 1{]}	    & $2.25\times10^{-6}$ \\
		\hline
	\end{tabular}
\end{table}

\subsection{Advection equation}
\label{sec:sec4.4}
\paragraph{}
Advection is one of the most significant processes in atmospheric motion, which is basically described by Advection equation. Here, we consider a 1-D Advection equation:

\begin{equation}
\label{equ:equ15}
	\begin{array}{l}
{u_t} + \beta {u_x} = 0,{\rm{                            }}x \in \left( {0,1} \right),t \in \left( {0,2} \right]\\
u(0,x) = 0.8\sin \left( {4\pi x + \pi /4} \right)
\end{array}
\end{equation}

Where $\beta$ is the advection speed and this equation has an analytical solution \cite{M2022PDEBench}:

\begin{equation}
\label{equ:equ16}
	u(t,x) = 0.8\sin \left[ {4\pi \left( {x - \beta t} \right) + \pi /4} \right]
\end{equation}

\begin{figure}
	\centering
	\includegraphics[width=17cm]{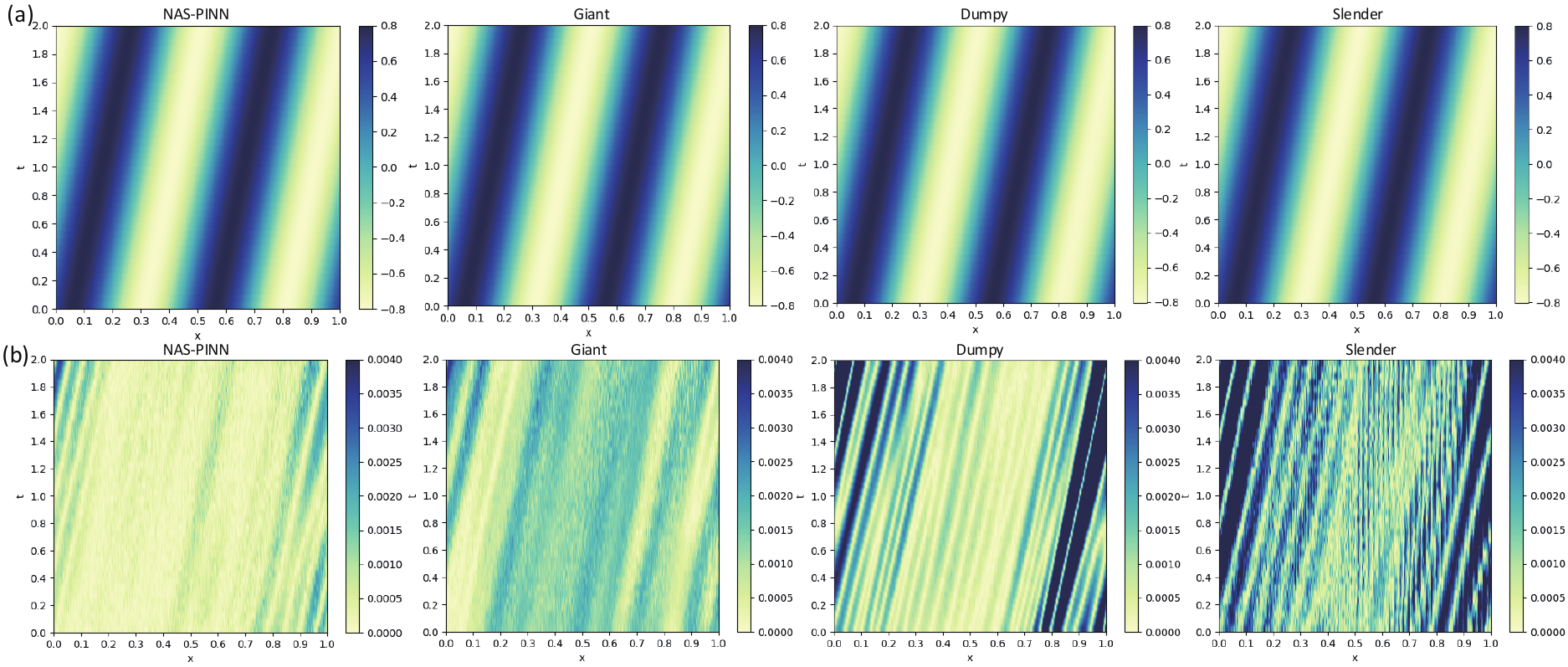}
	\caption{Advection equation: The predicted solutions (a) and the error distributions (b) of different neural architectures ($\beta=0.1$).}
	\label{fig:fig10}
\end{figure}

\begin{figure}
	\centering
	\includegraphics[width=17cm]{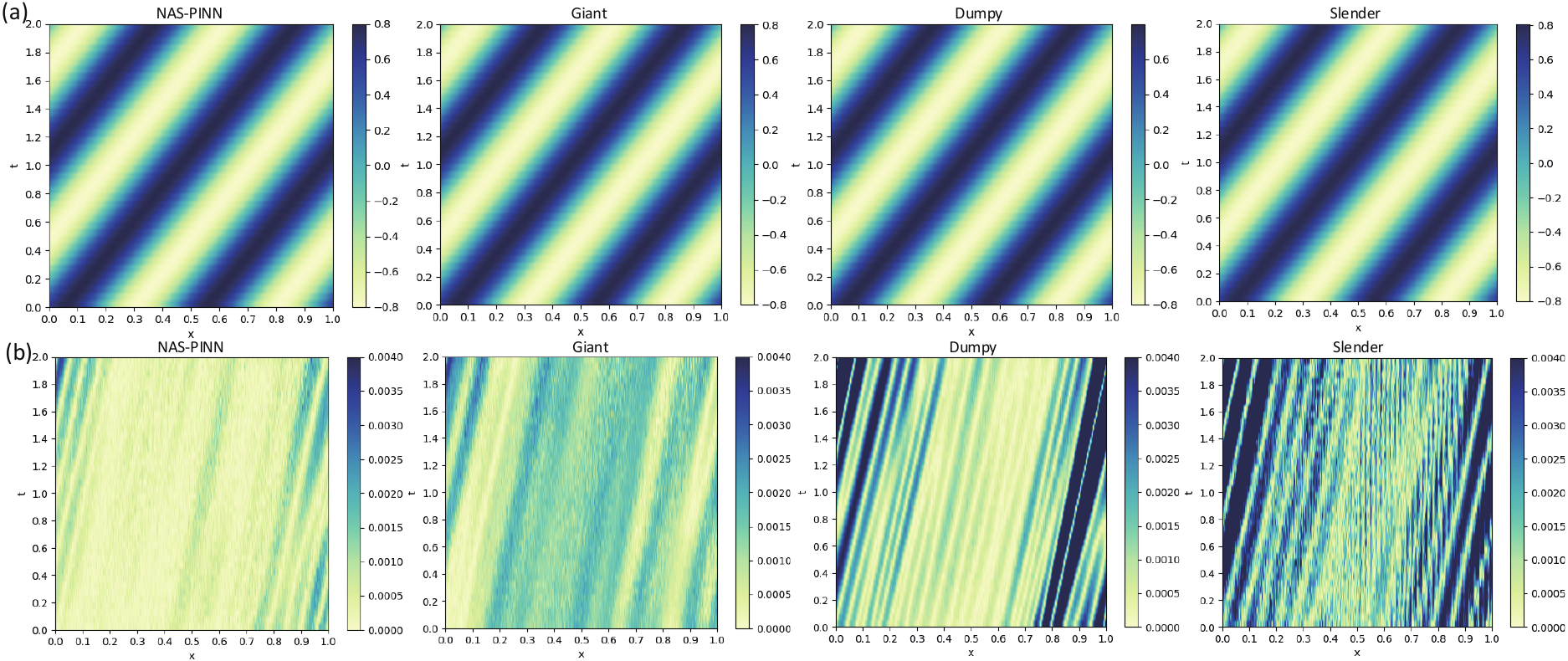}
	\caption{Advection equation: The predicted solutions (a) and the error distributions (b) of different neural architectures ($\beta=0.4$).}
	\label{fig:fig11}
\end{figure}

\paragraph{}
The Advection equations with different values of $\beta$ are investigated. The search space and the reference neural architectures are the same as above. 40 points along the $t$-axis and 120 points along the $x$-axis are uniformly taken for the architecture search phase. The same points are used to train all the neural architectures from scratch. 40 points along the t-axis and 120 points along the x-axis are uniformly sampled to test all the trained models. Table \ref{tab:tab4} and Figures \ref{fig:fig10} and \ref{fig:fig11} show the results of Advection equations with different advection speeds $\beta$, and the average values are taken from 5 independent experiments.

\paragraph{}
The NAS-PINN-searched neural architectures can always achieve the best results and it shows an obvious pattern that the smaller the $\beta$ is, the deeper the neural network will be. This pattern still works when we consider about the reference architectures for that the architecture Dumpy performs best when $\beta=1$ and the architecture Giant performs best when $\beta=0.1$. The architecture Slender almost fails when $\beta=1$, but the results become better with the decrease of $\beta$, and when $\beta=0.1$, the architecture Slender can even get equivalent results compared to the other two reference architectures. Similarly, the architecture Decrease achieves the second best accuracy when $\beta=0.1$, but gets the second worst results when $\beta=1$. Such results reveal that a same neural architecture may get quite different effects when faced with different PDEs or conditions, which can be a trouble to manually designed neural architectures and emphasizes the significance of research on NAS for PINNs. Meanwhile, the $L^2$ error indicates that the equation is more difficult to solve when the $\beta$ is larger. That reminds us that a complicated problem does not always need a deep neural network while a relatively simple problem may require more parameters to solve, and such phenomenon further emphasizes the value of NAS-PINN. Besides, similar to the results of Burgers equations, when a relatively deep neural network is constructed, different numbers of neurons are preferred.

\begin{table}[htbp]
	\centering
	\caption{Advection equation: $L^2$ error with different advection speeds $\beta$}
	\label{tab:tab4}
	\begin{tabular}{|l|l|l|l|}
		\hline
		$\beta$                      & Name         & Architecture                        & $L^2$ error  \\
		\hline
		\multirow{6}{*}{$\beta=1$}   & NAS-PINN             & {[}2, 110, 110, 1{]}                & $1.49\times10^{-4}$ \\
		\cline{2-4}
                       		& Giant   & {[}2, $110\times7$, 1{]}                   & $6.08\times10^{-4}$ \\
		\cline{2-4}
                       		& Dumpy   & {[}2, $110\times2$, 1{]}                   & $1.49\times10^{-4}$ \\
		\cline{2-4}
                       		& Slender & {[}2, $10\times7$, 1{]}                    & $2.67\times10^{-2}$ \\
		\cline{2-4}			
							& Increase & {[}2, 10, 30, 50, 70, 90, 110, 110, 1{]}	    & $8.56\times10^{-4}$ \\
		\cline{2-4}
							& Decrease & {[}2, 110, 90, 70, 50, 30, 10, 10, 1{]}	    & $1.40\times10^{-3}$ \\
		\hline
		\multirow{6}{*}{$\beta=0.4$} & NAS-PINN             & {[}2, 90, 90, 90, 90, 110, 1{]}     & $1.30\times10^{-6}$ \\
		\cline{2-4}
                       		& Giant   & {[}2, $110\times7$, 1{]}                   & $3.63\times10^{-6}$ \\
		\cline{2-4}
                       		& Dumpy   & {[}2, $110\times2$, 1{]}                   & $1.50\times10^{-6}$ \\
		\cline{2-4}
                       		& Slender & {[}2, $10\times7$, 1{]}                    & $3.61\times10^{-5}$ \\
		\cline{2-4}			
							& Increase & {[}2, 10, 30, 50, 70, 90, 110, 110, 1{]}	    & $5.25\times10^{-6}$ \\
		\cline{2-4}
							& Decrease & {[}2, 110, 90, 70, 50, 30, 10, 10, 1{]}	    & $4.90\times10^{-6}$ \\
		\hline
		\multirow{6}{*}{$\beta=0.1$} & NAS-PINN             & {[}2, 110, 50, 50, 70, 30, 90, 1{]} & $2.56\times10^{-6}$ \\
		\cline{2-4}
                       		& Giant   & {[}2, $110\times7$, 1{]}                   & $6.14\times10^{-6}$ \\
		\cline{2-4}
                       		& Dumpy   & {[}2, $110\times2$, 1{]}                   & $6.37\times10^{-6}$ \\
		\cline{2-4}
                       		& Slender & {[}2, $10\times7$, 1{]}                    & $8.55\times10^{-6}$ \\
		\cline{2-4}			
							& Increase & {[}2, 10, 30, 50, 70, 90, 110, 110, 1{]}	    & $3.31\times10^{-6}$ \\
		\cline{2-4}
							& Decrease & {[}2, 110, 90, 70, 50, 30, 10, 10, 1{]}	    & $2.77\times10^{-6}$ \\
		\hline
	\end{tabular}
\end{table}

\subsection{2-D Burgers equation}
\label{sec:sec4.5}
\paragraph{}
To explore the performance and characteristics of high dimensional problems, we consider a time-varying 2-D Burgers equation \cite{Y2022DeLISA}. As the network takes $(t, \bf{x})$ together as input, it can be reckoned as a 3-D problem here.

\begin{equation}
\label{equ:equ17}
	\begin{array}{l}
{u_t} + u({u_x} + {u_y}) = 0.1({u_{xx}} + {u_{yy}}),{\rm{         }}(x,y) \in \left[ {0,1} \right] \times \left[ {0,1} \right],t \in \left[ {0,2} \right]\\
u(0,x,y) = \frac{1}{{1 + \exp \left( {\frac{{x + y}}{{0.2}}} \right)}}\\
u(t,{x_b},{y_b}) = \frac{1}{{1 + \exp \left( {\frac{{{x_b} + {y_b} - t}}{{0.2}}} \right)}}
\end{array}
\end{equation}

\paragraph{}
This equation can be analytically solved:

\begin{equation}
\label{equ:equ18}
	u(t,x,y) = \frac{1}{{1 + \exp \left( {\frac{{x + y - t}}{{0.2}}} \right)}}
\end{equation}

\begin{figure}
	\centering
	\includegraphics[width=17cm]{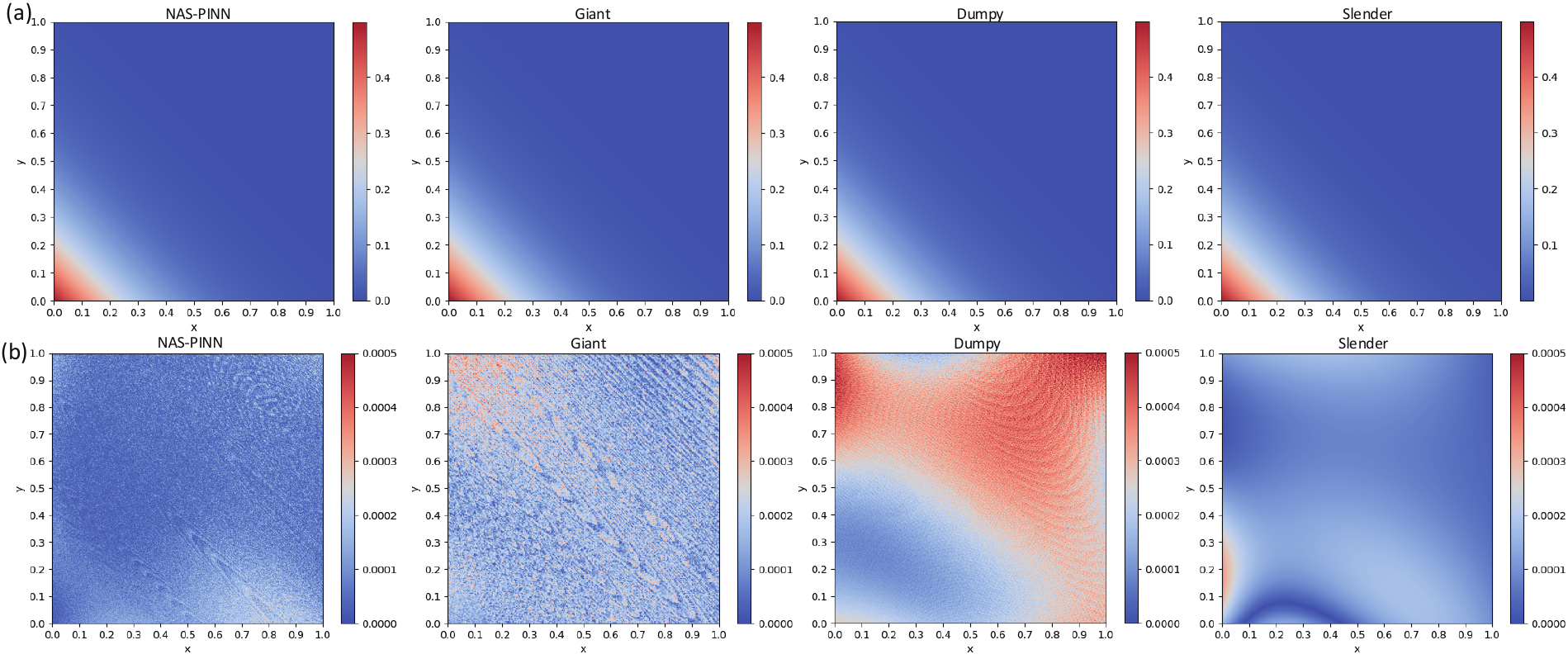}
	\caption{2-D Burgers equation: The predicted solutions (a) and error distributions (b) of different neural architectures ($t=0$).}
	\label{fig:fig12}
\end{figure}

\begin{figure}
	\centering
	\includegraphics[width=17cm]{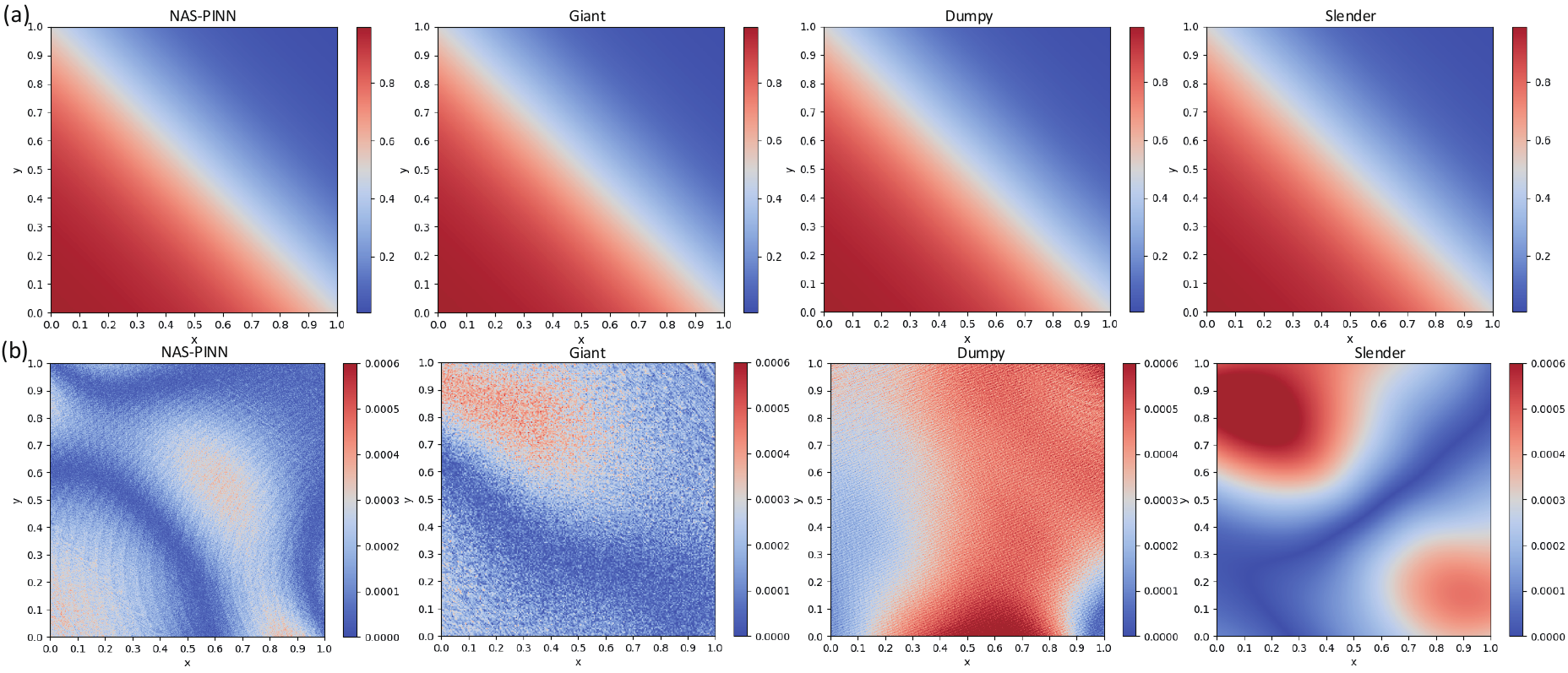}
	\caption{2-D Burgers equation: The predicted solutions (a) and error distributions (b) of different neural architectures ($t=1$).}
	\label{fig:fig13}
\end{figure}

\begin{figure}
	\centering
	\includegraphics[width=17cm]{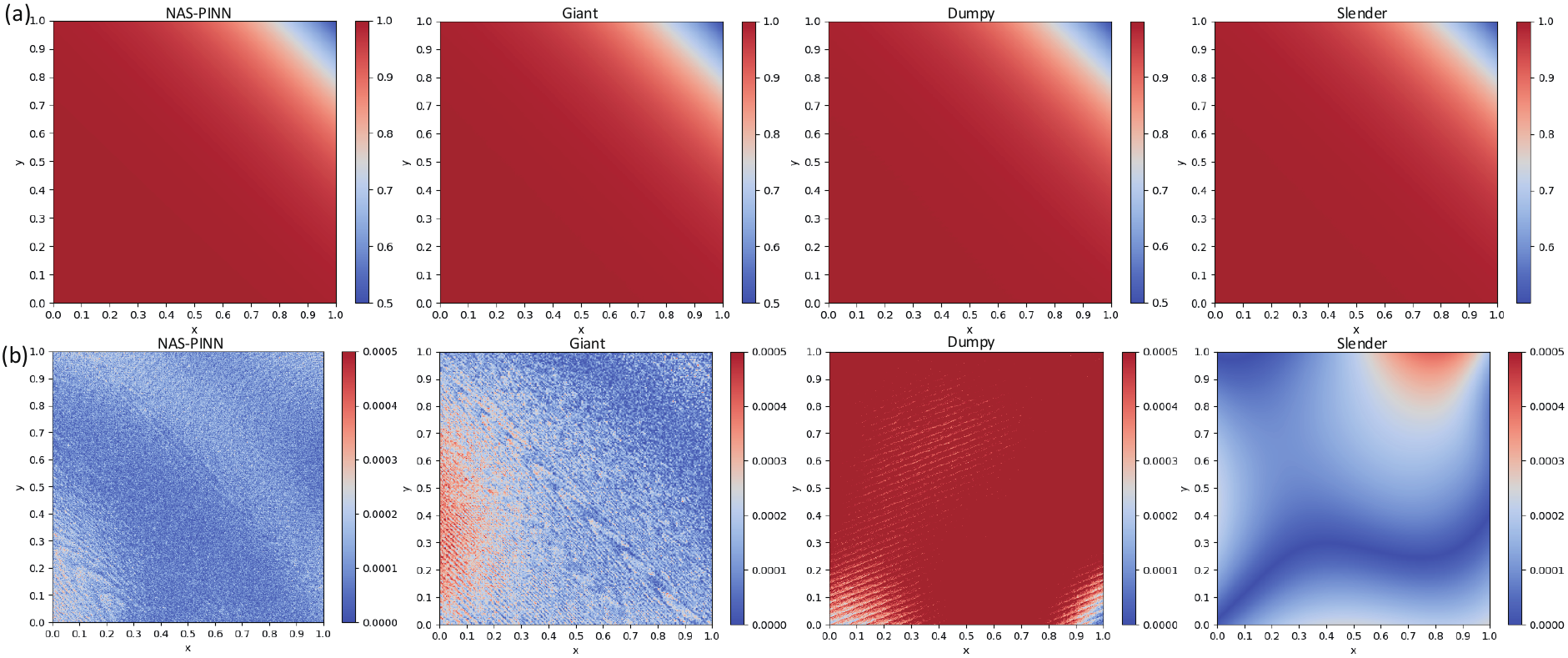}
	\caption{2-D Burgers equation: The predicted solutions (a) and error distributions (b) of different neural architectures ($t=2$).}
	\label{fig:fig14}
\end{figure}

\paragraph{}
The search space and the reference neural architectures are the same as above. In both the architecture search phase and the training phase, we uniformly take 20 points along the $t$-axis and 25 points along the $x$-axis and $y$-axis. To test the trained models, 41 points along the t-axis and 500 points along the x-axis and y-axis are uniformly sampled. Table \ref{tab:tab5} lists the corresponding $L^2$ error and Figures \ref{fig:fig12}-\ref{fig:fig14} show the results of 2D Burgers equation in three time slices, $t$=0, 1 and 2, respectively. Still, the average values are taken from 5 independent experiments. The architecture obtained by NAS-PINN achieves the minima $L^2$ error and the best error distributions. Compared to the architectures Dumpy, Slender, Increase and Decrease, the NAS-PINN-searched architecture can raise the accuracy by an order of magnitude. Although the Architecture Giant achieves an accuracy close to that of NAS-PINN, the parameters of Giant are obviously much more than NAS-PINN. The results show that the proposed NAS-PINN can adapt well to three dimensional problems. Similar to the conclusions of Burgers equations in Section \ref{sec:sec4.3}, having more layers is more crucial for getting better performance than having more neurons. However, there exists a most suitable number of hidden layers and too many layers can make the performance deteriorate. Again, it is recommended to vary the number of neurons used rather than sticking to the conventional wisdom of maintaining the number of neurons constant.

\begin{table}[htbp]
	\centering
	\caption{2-D Burgers equation: $L^2$ error of different neural architectures}
	\label{tab:tab5}
	\begin{tabular}{|l|l|l|}
		\hline
		Name         & Architecture                   & $L^2$ error \\
		\hline
		NAS-PINN             & {[}3, 110, 90, 70,   110, 1{]} & $4.29\times10^{-8}$  \\
		\hline
		Giant   & {[}3, $110\times7$, 1{]}              & $8.52\times10^{-8}$  \\
		\hline
		Dumpy   & {[}3, $110\times2$, 1{]}              & $1.61\times10^{-7}$  \\
		\hline
		Slender & {[}3, $10\times7$, 1{]}               & $1.53\times10^{-7}$  \\
		\hline			
		Increase & {[}2, 10, 30, 50, 70, 90, 110, 110, 1{]}	    & $2.22\times10^{-7}$ \\
		\hline
		Decrease & {[}2, 110, 90, 70, 50, 30, 10, 10, 1{]}	    & $2.73\times10^{-7}$ \\
		\hline
	\end{tabular}
\end{table}

\section{CONCLUSIONS}
\label{sec:sec5}
\paragraph{}
In this paper, we propose a neural architecture search-guided method, namely NAS-PINN, to automatically search the best neural architecture for solving given PDEs. By constructing the mixed operation and introducing the masks to realize the addition of tensors in different shapes, the architecture search problem can be relaxed into a continuous, bi-level optimization problem. It can search for the most suitable number of hidden layers and number of neurons for each layer in the given search space, and construct the best neural architecture for a given problem.

\paragraph{}
Through various numerical experiments, we verify the effectiveness of NAS-PINN and show its strong adaptability to irregular computational domains and high dimensional problems. A comparison between NAS-PINN and SMAC is also made, and the results demonstrate that NAS-PINN can search in a larger and more flexible search space with higher efficiency than SMAC. The reported numerical results further prove that more hidden layers do not necessarily mean better performance and sometimes more hidden layers can even be harmful. For Poisson equations and Advection equations, a relatively shallow neural network with more neurons in each layer can prevail over a deep one, which is quite against our common sense. Regardless of the input dimension, to have more layers appears to be crucial to solving Burgers equations, even when the number of neurons is fairly small. The numerical experiments also indicate that a neural network with different numbers of neurons for each layer is preferable to one whose hidden layers all have the same number of neurons. Another important fact is that effective neural architectures to solve different PDEs can drastically vary from each other. Such diversity can bring great trouble to the manual design of neural architecture, while NAS-PINN can help investigate characteristics of effective neural architectures and then improve research efficiency. Furthermore, the proposed method can be easily applied to other PDEs and to exploiting features of efficient neural architectures for solving those equations.

\paragraph{}
NAS-PINN has focused on DNNs so far, and based on the framework of NAS-PINN, the search for convolutional neural networks (CNNs) can be realized in the future. A selection between DNN and CNN can be made automatically as well. Besides, the threshold of whether to reserve mixed layers is worthy of further study. Besides, the threshold of whether to reserve mixed layers and how to improve the computing efficiency of mixed models from the algorithm level are worthy of further study.

\section*{Acknowledgments}
\label{sec:acknowledgments}
\paragraph{}
This work was supported in part by the National Natural Science Foundation of China (92066106), the Young Scientific and Technical Talents Promotion Project of Jiangsu Association for Science and Technology (2021031), the Zhishan Young Scholar Project of Southeast University (2242022R40022), and the Fundamental Research Funds for the Central Universities (2242022R40022).

\bibliographystyle{unsrt}


\end{document}